\documentclass[fleqn,twoside]{article}
\usepackage{espcrc2}

\usepackage{graphicx}
\usepackage[figuresright]{rotating}

\def\SU#1{{\rm SU(#1)}}
\def\SO#1{{\rm SO(#1)}}
\def\U#1{{\rm U(#1)}}
\def\ii{{\rm i}}
\def\slash#1{\rlap{\kern -.2bp/}#1}
\newcommand{\e}{{\rm e}}
\newcommand{\muegamma}{\mu\to\e\gamma}
\newcommand{\mueee}{\mu\to\e\e\overline{\e}}
\newcommand{\mueN}{\mu\, {\rm N\to\e\, N}}
\newcommand{\bea}{\begin{eqnarray}}
\newcommand{\eea}{\end{eqnarray}}
\newcommand{\ba}{\begin{array}}
\newcommand{\ea}{\end{array}}

\newcommand{\AmS}{{\protect\the\textfont2
  A\kern-.1667em\lower.5ex\hbox{M}\kern-.125emS}}

\title{One-loop flavor change in Little Higgs models\thanks{Talk given by J.I.I. at ``Loops and Legs in Quantum Field Theory", 10th DESY Workshop on Elementary Particle Theory, 25-30 April 2010, W\"orlitz, Germany.}}

\author{Jos\'e I. Illana\address[ugr]{%
        CAFPE and Departamento de F{\'\i}sica Te\'orica y del Cosmos, \\ 
        Universidad de Granada, E-18071 Granada, Spain}
        and Mark D. Jenkins\addressmark[ugr]}
       
\begin{document}

\begin{abstract}
The Little Higgs (LH) idea attempts to cure the little amount of fine-tuning
necessary to bridge the gap between the Higgs mass (electroweak scale) and the new physics scale suggested by electroweak precision tests ($\sim10$~TeV). However, we show that LH models do not survive the confrontation with experimental limits on lepton flavor mixing, assuming the same naturalness arguments that motivate their introduction. Two different LH models are analyzed and several aspects of their one-loop predictions for lepton flavor-changing processes are discussed.

\vspace{1pc}
\end{abstract}

\maketitle

\setcounter{footnote}{0}
\section{The hierarchy and the flavor problems}

The Standard Model (SM) suffers a hierarchy problem: the Higgs mass must be of the order of the electroweak scale in spite of receiving quadratic loop corrections of the order of the Planck scale (theory cutoff), when gravity sets in. Naturalness arguments then require that some new physics must show up earlier, around the TeV, just the region to be explored by the LHC. 

Supersymmetry is one of the favourite scenarios for physics beyond the SM. In supersymmetric models extra loop contributions from superpartners approximately cancel the quadratic Higgs mass corrections, provided that the supersymmetry-breaking terms are at most of the order of the TeV. However, a {\em little hierarchy} problem persists because in most of the parameter space of these theories the Higgs mass lies below the current experimental lower bound, implying a fine-tuning at the few percent level \cite{Birkedal:2004xi}.

The Little Higgs (LH) models \cite{ArkaniHamed:2001ca} were introduced to offer an explanation to the little hierarchy between the electroweak scale $v=246$~GeV and the expected new physics scale $f\sim 1$~TeV. In these models the Higgs is a pseudo-Goldstone boson of an approximate global symmetry broken at $f$. The physics below a cutoff $\Lambda\sim 4\pi f \sim 10$~TeV, is effectively described by a non-linear sigma model. For physics above $\Lambda$, ultraviolet completion of the theory would be required, that is both unknown and beyond the current experimental reach. The global symmetry is explicitly broken by gauge and Yukawa interactions, giving the Higgs a mass and non-derivative interactions while preserving the cancellation of just the one-loop quadratic corrections, thanks to the collective symmetry breaking \cite{LH-reviews}. The sensitivity to a 10~TeV cutoff at two loops is not considered unnatural.

On the other hand, neither supersymmetric nor LH models are taylored to explain the observed suppression of flavor-changing neutral currents (FCNC), that can be naturally accommodated in the SM by the GIM mechanism but demands alignment between the standard Yukawa matrices and the new physics interactions. It can be argued that supergravity provides the initial conditions in the supersymmetry framework to explain the needed fine tuning, but in LH models the introduction of new gauge bosons and fermions with arbitrary flavor structure challenges the flavor problem.

The purpose of this talk is to report on recent works \cite{delAguila:2008zu,paper2,paper3} that have explored the naturalness of two different LH models as they confront the experimental limits on lepton flavor-changing processes. Previous complementary studies on flavor violation focused on a single model, the Littlest Higgs without \cite{Buras:2006wk} and with T-parity, for the quark \cite{Blanke:2006eb,Goto:2008fj} and the lepton \cite{Choudhury:2006sq,Blanke:2007db} sectors.

\section{Lepton flavor mixing in Little Higgs}

There are essentially two types of LH models: the Product group models, where the SM $\SU{2}_L$ group comes from the diagonal breaking of two or more gauge groups, and the Simple group models, where the $\SU{2}_L$ comes from the breaking of a larger group. We analyze below an example of each type.

\subsection{Littlest Higgs with T-parity}

The Littlest Higgs with T-parity (LHT) \cite{ArkaniHamed:2002qy,Cheng:2003ju,Low:2004xc} is a non-linear sigma model based on the breaking at a scale $f$ of a global $\SU{5}\to\SO{5}$ by the VEV of a $5\times5$ symmetric tensor. The broken generators expand 14 Goldstone fields. A subgroup $[\SU{2}\times\U{1}]_1\times[\SU{2}\times\U{1}]_2\subset\SU{5}$ is gauged and gets broken to the SM $\SU{2}_L\times\U{1}_Y$. The gauge bosons associated to the 4 unbroken generators ($W^\pm,Z,\gamma$) will remain massless and the other 4 ($W_H^\pm,Z_H,A_H$) get masses of the order of $f$ after eating the corresponding would-be-Goldstone bosons. In a second step, the electroweak symmetry breaking (EWSB) $\SU{2}_L\times\U{1}_Y\to \U{1}_{\rm QED}$ takes place when a complex $\SU{2}$ doublet (4 of the 10 remaining Goldstone bosons) is identified with the SM Higgs (3 would-be-Goldstone bosons, to be eaten by $W^\pm,Z$, plus the physical $h$). In addition, a discrete symmetry (T-parity) is introduced under which the SM fields are even and (most of) the new ones are odd, in order to be consistent with electroweak precision data \cite{Hubisz:2005tx,Chen:2006dy}.

The gauge and T-parity invariant Lagrangian of the gauge and scalar sectors is easy to obtain in terms of covariant derivatives \cite{Blanke:2006eb}. The fermion sector is much more complicated. For each generation, one  has a SM (extra) left-handed fermion doublet from a T-even (T-odd) combination of two incomplete representations under $\SU{5}$, $\Psi_1$[${\bf\bar5}$] and $\Psi_2$[${\bf5}$]. In addition, an extra right-handed doublet is introduced in a non-linear five-dimensional representation $\Psi_R$ of $\SO{5}$.\footnote{Actually, $\xi\Psi_R$ is a ${\bf5}$ under $\SU{5}$, where $\xi=\exp\{\ii\Pi/f\}$ is the exponential of the Goldstone fields.} $\Psi_1$ and $\Psi_2$ are connected by a T-transformation whereas $\Psi_R$ is a T-eigenstate. A Yukawa Lagrangian that couples $\Psi_1$ and $\Psi_2$ to $\Psi_R$ provides heavy masses of order $f$ to the extra fermion doublet, respecting gauge an T symmetries. Then the gauge interactions of the light left-handed fermions to the heavy ones are fixed and involve the Goldstone fields \cite{Hubisz:2004ft}, introducing corrections to the vertices coupling two heavy right-handed fermions to a SM gauge boson \cite{delAguila:2008zu,Goto:2008fj} that are crucial to keep the one-loop amplitudes ultraviolet finite. Another subtlety concerns the SM right-handed singlets, whose hypercharges can only be accommodated by enlarging $\SU{5}$ with two extra $\U{1}$ factors \cite{Goto:2008fj}.

In the case of three generations, the interaction of a heavy gauge boson with a light and a heavy left-handed fermion is proportional to the misalignment of the respective Yukawa matrices:
\bea
{\rm quarks:}  && V_{Hu}=V_H^{q\dagger} V_u\ , \quad
                  V_{Hd}=V_H^{q\dagger} V_d\ , \\
{\rm leptons:} && V_{H\nu} =V_H^{l\dagger} V_\nu \ , \quad\
                  V_{H\ell}=V_H^{l\dagger} V_\ell\ ,
\eea
where $V_{\rm CKM}=V_u^\dagger V_d$ and $V^q_H$ is the rotation performed to heavy left-handed quarks to diagonalize the Yukawa matrix of the heavy fermions (analogously, $V_{\rm PMNS}=V_\nu V_\ell^\dagger$ in the lepton sector). One has in particular lepton-flavor changing interactions of charged and neutral currents proportional to
$V_{H\ell}^{i\alpha}\bar\nu_{HL}^i\slash{W}_H^\dagger\ell_L^\alpha$,
$V_{H\ell}^{i\alpha}\bar\ell_{HL}^i\slash{A}_H^\dagger\ell_L^\alpha$ and
$V_{H\ell}^{i\alpha}\bar\ell_{HL}^i\slash{Z}_H^\dagger\ell_L^\alpha$.

\subsection{Simplest Little Higgs}

The Simplest Little Higgs (SLH) \cite{Schmaltz:2004de} is based on the breaking of the global symmetry $G\equiv[\SU{3}\times\U{1}]_1\times[\SU{3}\times\U{1}]_2\to[\SU{2}\times\U{1}]_1\times[\SU{2}\times\U{1}]_2$ when two scalar triplets $\Phi_1[({\bf 3},{\bf 1})]$ and $\Phi_2[({\bf 1},{\bf 3})]\subset G$ acquire VEVs $f\cos\beta$ and $f\sin\beta$, respectively. The broken generators expand 10 Goldstone fields. A subgroup $\SU{3}\times\U{1}_\chi$ is gauged and gets broken to the SM group. Then 4 gauge bosons remain massless ($W^\pm,Z,\gamma$) and 5 get heavy masses of order $f$ ($X^\pm,Y^0,\bar Y^0,Z'$) after eating 5 would-be-Goldstone bosons. The other 5 Goldstones include the SM Higgs doublet, that provides masses to the weak bosons after the EWSB at a lower scale.

Again it is straightforward to express the Lagrangian for the scalar and gauge sectors in terms of covariant derivatives. The would-be-Goldstone fields in the 't Hooft-Feynman gauge are also obtained after a good amount of algebra \cite{paper3}.
To build the fermion sector, the SM fermions are included into $\SU{3}$ multiplets together with a new heavy fermion.  Each lepton family $i$ consists of an $\SU{3}$ left-handed triplet [${\bf3}$], $(\nu_L,\ell_L,\ii N_L)_i$, and two right-handed singlets, $\ell_{Ri}$ and $N_{Ri}$. For quarks there are two options. In the {\em universal embedding} the quarks are introduced similarly to the leptons: $(u_L,d_L,\ii U_L)_i$, $u_{Ri}$, $d_{Ri}$ and $U_{Ri}$. However this embedding is not anomaly free. Although this is not necessarily a problem in an effective theory, because additional fermions could be added at the cutoff scale to cancel the anomalies, one can construct an {\em anomaly-free embedding} with no additional degrees of freedom \cite{Kong:2003tf}: the third family is analogous to the lepton sector, $(t_L,b_L,\ii T_L)$, $t_R$, $b_R$, $T_R$, but the first two families contain conjugate triplets [${\bf\bar3}$], $(d_L,-u_L,\ii D_L)$, $d_R$, $u_R$, $D_R$ and $(s_L,-c_L,\ii S_L)$, $s_R$, $c_R$, $S_R$, respectively.

The flavor mixing in the quark sector is extremely rich. In general, all quarks
mix with other heavy and light quarks of every family. However we will be interested just in the effects that are consequence of the mixing in the lepton sector, so quark intergeneration mixing will be neglected \cite{Han:2005ru}. Concerning leptons, after the EWSB the light and the heavy neutrino of the same family mix at order $v/f$ and there is family mixing as long as the Yukawa matrix of heavy neutrinos and that of charged leptons are not aligned. Thus, in the basis where the heavy neutrinos are diagonal:
\bea
\ell^0_{Li}\!\!\!\!&=&\!\!\!\![V_\ell\ell_L]_i\ , \\
\left[\ba{c}\nu_{Li}^0\\N_{Li}^0\ea\right]\!\!\!\!&=&\!\!\!\!\left[\ba{cc}1 & -\delta_\nu \\ \delta_\nu & 1\ea\right]\left[\ba{c}[V_\ell\nu_L]_i\\N_{Li}\ea\right]+{\cal O}(\delta_\nu^2)\,,
\eea
with $\delta_\nu=-v/(\sqrt{2}f\tan\beta)$. This induces lepton-flavor changing interactions of charged currents proportional to
$V_\ell^{i\alpha}\overline N_{Li}\slash{X}^\dagger\ell_{L\alpha}$ and
$\delta_\nu V_\ell^{i\alpha}\overline N_{Li}\slash{W}^\dagger\ell_{L\alpha}$. Although the latter is suppressed by $v/f$, both types contribute to the amplitudes at the leading order.

\section{Lepton flavor-changing processes}

After expanding the Lagrangians of both models to the required order in $v/f$ one obtains the complete set of Feynman rules needed \cite{delAguila:2008zu,paper3}. We study all lepton flavor-changing processes available: $\muegamma$, $\mueee$ and $\mueN$ ($\mu$ to $\e$ conversion in nuclei). The contributions to all them vanish at tree level and those at one-loop are organized in form factors written in terms of generic couplings for the most general vertices involving scalar, fermion and vector fields. The invariant amplitudes are expressed as combinations of standard loop integrals \cite{Passarino:1978jh}, including bubbles, triangles and boxes, that have been computed algebraically and reduced to a simple form.

The leading order of the amplitudes is $v^2/f^2$. In the case of the LHT model, the T-parity forces only heavy particles (T-odd) in the loop. As a consequence, the loop integrals have a well defined order. In constrast, in the SLH model there are heavy and light neutrinos or gauge bosons in the same loop, that demands a careful $v/f$ expansion of the loop functions for consistency, since the coupling constants in the Feynman rules are obtained to a given order \cite{paper3}.

As an important remark, all form factors are ultraviolet finite in both models. Box contributions are finite by simple power counting. The dipole part of the penguins is finite even for generic couplings and it is subleading except for photon penguins. The chirality-conserving (non-dipole) part of the penguins is left-handed and receives contributions from triangle and flavor-changing bubble diagrams, with divergences that cancel exactly thanks to the unitarity of the flavor mixing matrices. Incidentally, in the LHT model the $Z$-penguin receives contributions only from the $W_H$, the heavy partner of the $W$ (those from the $A_H$ and $Z_H$ vanish), and its form factor has exactly the same appearance as that of the SM with massive neutrinos \cite{Illana:2000ic}.

\begin{figure*}[p]
\begin{center}
\renewcommand{\arraystretch}{6}
\begin{tabular}{ccc}
\vspace*{-5mm}
\includegraphics[width=.37\textwidth]{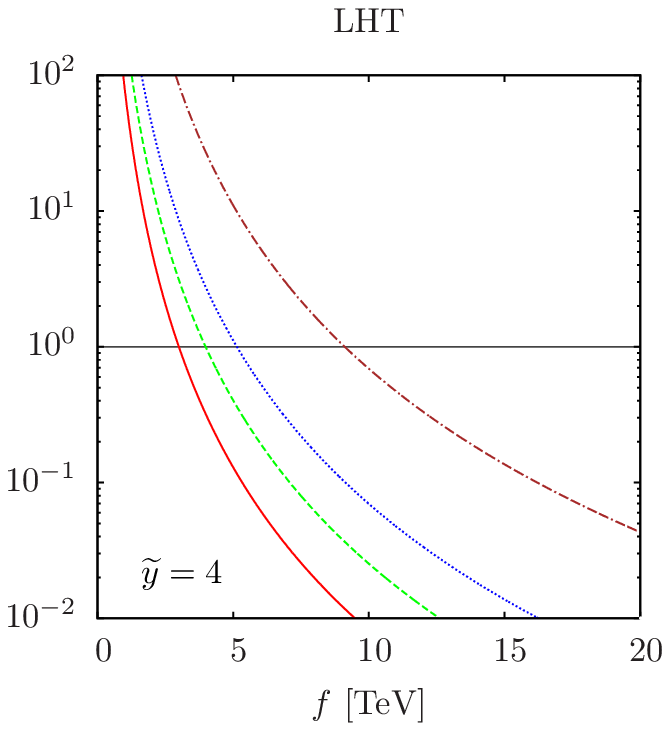}
&\qquad&
\includegraphics[width=.37\textwidth]{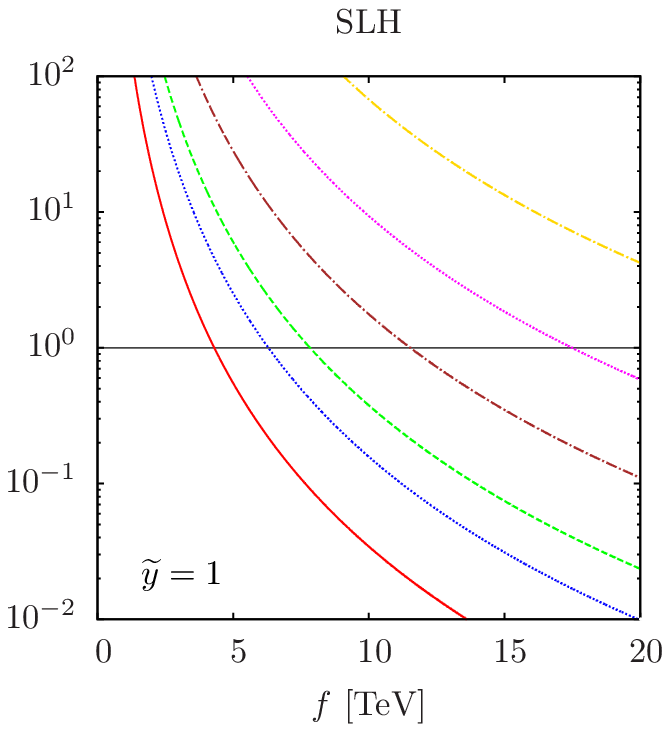}
\\
\vspace*{-5mm}
\includegraphics[width=.37\textwidth]{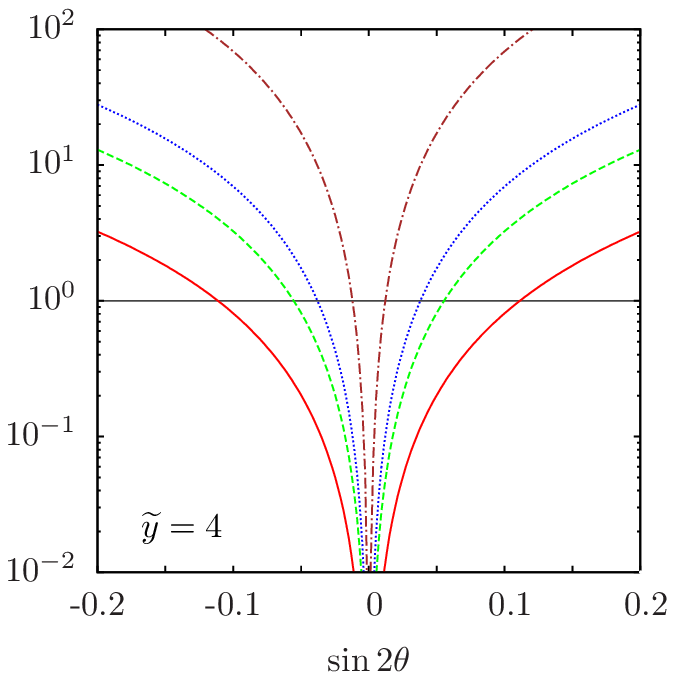}
&\qquad&
\includegraphics[width=.37\textwidth]{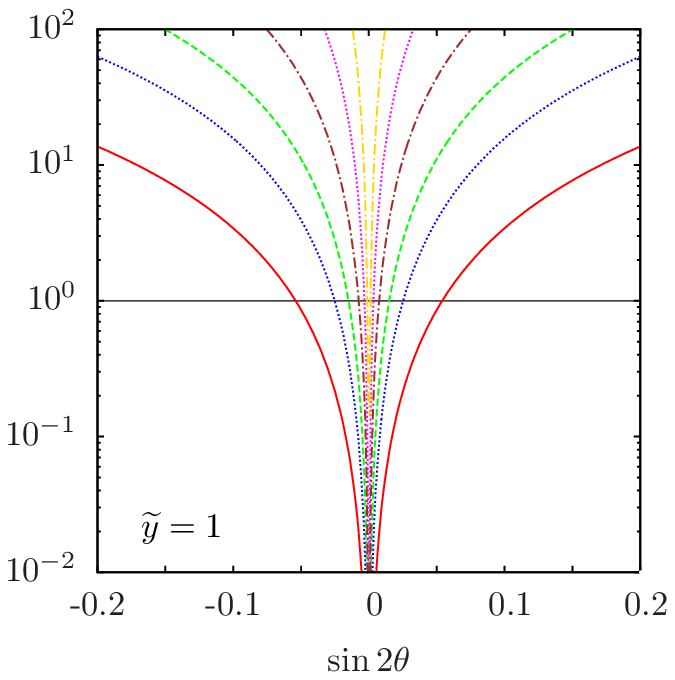}
\\
\vspace*{-5mm}
\includegraphics[width=.37\textwidth]{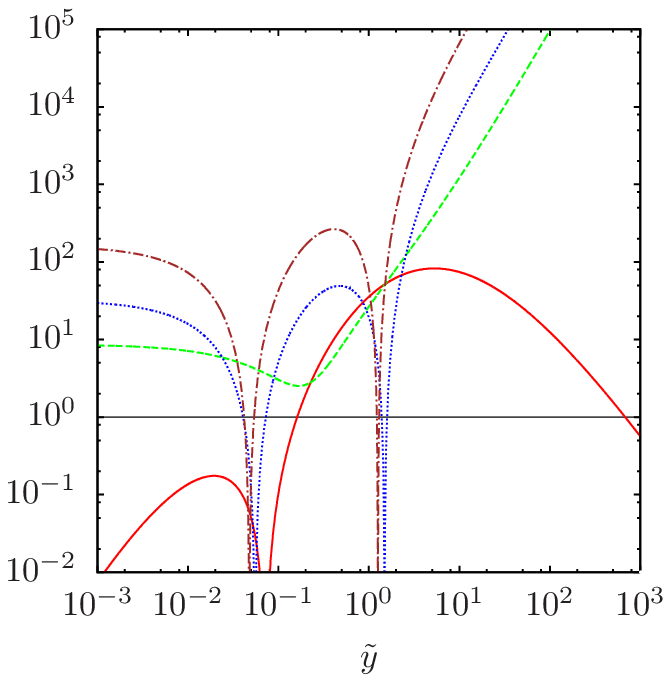}
&\qquad&
\includegraphics[width=.37\textwidth]{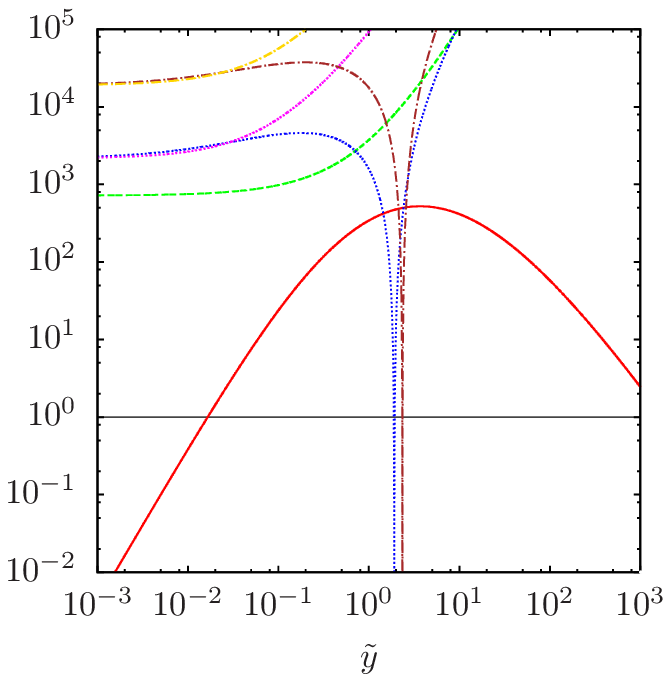}
\end{tabular}
\end{center}
\vspace*{-10mm}
\caption{Ratios of expectations to current limits for $\muegamma$ (solid), $\mueee$ (dashed), $\mu\;{\rm Ti}\to\e\;{\rm Ti}$ (dotted) and $\mu\;{\rm Au}\to\e\;{\rm Au}$ (dot-dashed) in the LHT (left) and the SLH (right). The $\mueN$ curves in the SLH model are doubled: the lower (upper) ones assume anomaly-free (universal) embedding.}
\label{fig1}
\end{figure*}

In order to simplify the phenomenological analysis, we assume just two lepton generations. Then the heavy lepton sector consists of $\nu_{Hi}$ and $\ell_{Hi}$ for the LHT model\footnote{A difference of order $v^2/f^2$ in the masses of $\nu_{Hi}$ and $\ell_{Hi}$ is neglected.} and $N_i$ for the SLH model, with $i=1,2$. The mixing matrix is parameterized by
\bea
V=\left[\ba{cc} V^{1e} & V^{1\mu} \\ V^{2e} & V^{2\mu} \ea\right]
 =\left[\ba{rr} \cos\theta & \sin\theta \\ -\sin\theta & \cos\theta \ea\right]\ .
\eea
The gauge boson masses are determined by the breaking scale $f$ and the heavy fermions masses can be defined as $m^2_{Hi}\equiv y_i M^2_{W_H}$ (LHT) or $m^2_{Hi}\equiv y_i M^2_X$ (SLH). We are left with just 4 free parameters. Our choice is
\bea
f\ ,\
\theta\ ,\
\delta=\frac{m_{H2}^2-m_{H1}^2}{m_{H1}m_{H2}}\ ,\
\widetilde y=\sqrt{y_1y_2}\ .
\eea
The amplitudes scale as $v^2/f^2$, are roughly proportional to $\sin2\theta$ and $\delta$ and vary with $\widetilde y$. Namely, the flavor violation effects are suppressed by the Little Higgs scale and vanish unless the heavy fermions mix and are non degenerate. {\em Naturalness} suggests that these parameters take the following {\em reference values}:
\bea
f\sim 1\mbox{ TeV}\ , \
\sin2\theta\sim 1\ , \
\delta\sim 1\ , \
\widetilde y\sim 1\ .
\eea
For the process $\mueN$ the heavy quarks also play a role. By default we will assume that they are degenerate and with masses of the order of $M_{W_H}$ ($M_X$) in the LHT (SLH) model.

Figure~\ref{fig1} shows our predictions for the ratios of branching fractions of $\muegamma$, $\mueee$ and the conversion rates of $\mueN$ (in gold and titanium) to the present experimental limits \cite{present} in the two Little Higgs models under study. One parameter is varied while the others are set to the reference values above. Ratios above unity are excluded. The curves as a function of $\delta$ are not displayed but are almost indistinguishable from those of $\sin2\theta$, which provides evidence for the  behaviour of the amplitudes pointed out above. Also the scaling with $f^{-4}$ of the ratios is obvious. The dependence on $\widetilde y$ is less evident and shows some interesting features. The observables involving $Z$ or $Z'$ penguins diverge in the limit of large $\widetilde y$ (notice that the curves for $\muegamma$ are the only well behaved). This is an example of non-decoupling similar to the case of the top quark in the $Zb\bar b$ vertex, that induces there corrections proportional to $m_t^2/M_W^2$. Furthermore, potential cancellations among the different contributions to the amplitudes depend critically on $\widetilde y$. The key observation is that the rates of the different processes do not simultaneously vanish for the same point of the parameter space, and therefore such regions of accidental suppressions do not really escape experimental constraint. That is why, to prevent an irrelevant cancellation in $\mueN$ a reference value $\widetilde y=4$, still of order one, has been chosen for the LHT model. All these plots are for degenerate quark masses. We have checked that in the LHT model, as long as we are far from a cancellation point, the quark mixing of non-degenerate quarks does not change the results much \cite{paper2}. As a final remark, $\mueN$ depends on $\beta$ in the SLH model \cite{paper3}. A safe reference value of $\tan\beta=1$ was taken along the study. Cancellations for other values are possible.

\section{Conclusions}

Tables~\ref{tab1} and \ref{tab2} summarize the present and future experimental constraints on the relevant parameters of the LHT and SLH models, based on naturalness arguments and avoiding accidental cancellations. Presently, $\mu\to\e$ conversion in Au is the most stringent process. LH models solve the little hierarchy problem if $f\sim1$~TeV. However, it is remarkable that current limits from lepton flavor-changing processes indicate a much heavier scale $f\sim 10$~TeV or else indicate a large amount of fine-tuning in the mixing with (or the mass splitting of) the heavy fermion sector.

\begin{table*}[t]
\caption{Bounds on LHT parameters from present \cite{present} and future$^*$ \cite{future} limits on lepton flavor-changing processes.}
\label{tab1}
\renewcommand{\tabcolsep}{.76pc} 
\renewcommand{\arraystretch}{1.2} 
\begin{tabular}{rccccccc}
\hline
\multicolumn{1}{c}{LHT} & \multicolumn{2}{c}{$\muegamma$} & 
  \multicolumn{2}{c}{$\mueee$}  &   $\mu\, {\rm Au}\to\e\, {\rm Au}$ &
  \multicolumn{2}{c}{$\mu\, {\rm Ti}\to\e\, {\rm Ti}$} \\
\hline
\multicolumn{1}{c}{Limit}
& $1.2\times10^{-11}$ & $10^{-13*}$ & $10^{-12}$ &  $10^{-14*}$ &  $7\times10^{-13}$ & $4.3\times10^{-12}$   &  $10^{-18*}$
\\
\hline
$f/\mbox{TeV}>$ & 3.00  & 9.93  & 3.98  & 12.6 & 9.11 & 5.13  & 234 \\
$\sin2\theta<$  & 0.111 & 0.010 & 0.055 & 0.006 & 0.012 & 0.038 & $<10^{-4}$ \\
$|\delta|<$     & 0.106 & 0.010 & 0.062 & 0.006 & 0.012 & 0.039 & $<10^{-4}$\\
\hline
\end{tabular}
\end{table*}

\begin{table*}[t]
\caption{Bounds on SLH parameters from present \cite{present} and future$^*$ \cite{future} limits on lepton flavor-changing processes. For $\mueN$ the numbers correspond to the anomaly-free (universal) embedding.}
\label{tab2}
\renewcommand{\tabcolsep}{.31pc} 
\renewcommand{\arraystretch}{1.2} 
\begin{tabular}{rcccccccccc}
\hline
\multicolumn{1}{c}{SLH} & \multicolumn{2}{c}{$\muegamma$} & 
  \multicolumn{2}{c}{$\mueee$}  &   \multicolumn{2}{c}{$\mu\, {\rm Au}\to\e\, {\rm Au}$} &
  \multicolumn{4}{c}{$\mu\, {\rm Ti}\to\e\, {\rm Ti}$} \\
\hline
\multicolumn{1}{c}{Limit}
& $1.2\times10^{-11}$ & $10^{-13*}$ & $10^{-12}$ &  $10^{-14*}$ &  \multicolumn{2}{c}{$7\times10^{-13}$} & \multicolumn{2}{c}{$4.3\times10^{-12}$}  &  \multicolumn{2}{c}{$10^{-18*}$}
\\
\hline
$f/\mbox{TeV}>$ & 4.3 & 14.2 & 7.8 & 24.8 & 11.5 & (28.7) & 6.3 & (17.5) & 287 & (796) \\
$\sin2\theta<$  & 0.0540 & 0.0048 & 0.0150 & 0.0014 & 0.0074 & (0.0012) & 0.0252 & (0.0032) & $<10^{-4}$ & ($<10^{-4}$) \\
$|\delta|<$     & 0.0517 & 0.0047 & 0.0159 & 0.0015 & 0.0070 & (0.0011) & 0.0232 & (0.0032) & $<10^{-4}$ & ($<10^{-4}$) \\
\hline
\end{tabular}
\end{table*}

\noindent{\em Acknowledgments.}
The authors are grateful to Paco del \'Aguila for a fruitful collaboration.  
Work supported by the Spanish MICINN (FPA2006-05294), Junta de Andaluc{\'\i}a (FQM 101, FQM 03048) and European Union (MRTN-CT-2006-035505: ``HEPTOOLS").

\end{document}